\documentclass[aps,prl,reprint,superscriptaddress,showpacs]{revtex4-1}

\usepackage{graphicx}

\newcommand{\aag}{$(\alpha,\alpha'\gamma)~$}

\newcommand{\krf}{$(\gamma,\gamma')~$}

\newcommand{\sn}{$^{124}$Sn}
\newcommand{\ba}{$^{138}$Ba}
\newcommand{\ce}{$^{140}$Ce}

\begin{document}

\title{Isospin character of the pygmy dipole resonance in \sn}

\author{J.~Endres}
\email[]{endres@ikp.uni-koeln.de}

\affiliation{Institut f\"ur Kernphysik, Universit\"at zu K\"oln, Z\"ulpicher Stra\ss{}e 77, D-50937 K\"oln, Germany}
\author{E.~Litvinova}
\affiliation{GSI Helmholtzzentrum f\"ur Schwerionenforschung, D-64291 Darmstadt, Germany}
\affiliation{Institut f\"ur Theoretische Physik, Goethe-Universit\"at, D-60438 Frankfurt am Main, Germany}
\author{D.~Savran}
\affiliation{Institut f\"ur Kernphysik, TU Darmstadt, D-64289 Darmstadt, Germany}
\author{P.A.~Butler}
\affiliation{Oliver Lodge Laboratory, University of Liverpool, L69 7ZE, United Kingdom}
\author{M.N.~Harakeh}
\affiliation{GSI Helmholtzzentrum f\"ur Schwerionenforschung, D-64291 Darmstadt, Germany}
\affiliation{Kernfysisch Versneller Instituut, University of Groningen, 9747 AA Groningen, The Netherlands}
\author{S.~Harissopulos}
\affiliation{Institute of Nuclear Physics, N.C.S.R. Demokritos Athens, GR-15310 Athens, Greece}
\author{R.-D.~Herzberg}
\affiliation{Oliver Lodge Laboratory, University of Liverpool, L69 7ZE, United Kingdom}
\author{R.~Kr\"ucken}
\affiliation{Physik Department, TU M\"unchen, D-85748 Garching, Germany}
\author{A.~Lagoyannis}
\affiliation{Institute of Nuclear Physics, N.C.S.R. Demokritos Athens, GR-15310 Athens, Greece}
\author{N.~Pietralla}
\affiliation{Institut f\"ur Kernphysik, TU Darmstadt, D-64289 Darmstadt, Germany}
\author{V.Yu.~Ponomarev}
\affiliation{Institut f\"ur Kernphysik, TU Darmstadt, D-64289 Darmstadt, Germany}
\author{L.~Popescu}
\affiliation{Kernfysisch Versneller Instituut, University of Groningen, 9747 AA Groningen, The Netherlands}
\affiliation{Belgian Nuclear Research Centre SCK*CEN, B-2400 Mol, Belgium}
\author{P.~Ring}
\affiliation{Physik Department, TU M\"unchen, D-85748 Garching, Germany}
\author{M.~Scheck}
\affiliation{Oliver Lodge Laboratory, University of Liverpool, L69 7ZE, United Kingdom}
\author{K.~Sonnabend}
\affiliation{Institut f\"ur Kernphysik, TU Darmstadt, D-64289 Darmstadt, Germany}
\author{V.I.~Stoica}
\author{H.J.~W\"ortche}
\affiliation{Kernfysisch Versneller Instituut, University of Groningen, 9747 AA Groningen, The Netherlands}
\author{A.~Zilges}
\affiliation{Institut f\"ur Kernphysik, Universit\"at zu K\"oln, Z\"ulpicher Stra\ss{}e 77, D-50937 K\"oln, Germany}

\date{\today}

\begin{abstract}
The pygmy dipole resonance has been studied in the proton-magic
nucleus $^{124}$Sn with the $(\alpha,\alpha'\gamma)$ coincidence
method at $E_{\alpha}=136~$MeV. The comparison with results of photon-scattering experiments reveals a splitting into two components with different structure: one group of
states which is excited in \aag as well as in \krf reactions and a group of
states at higher energies which is only excited in \krf reactions. Calculations
with the self-consistent relativistic quasiparticle time-blocking
approximation and the quasi-particle phonon model are in
qualitative agreement with the experimental results and predict a low-lying isoscalar component dominated by neutron-skin oscillations and a higher-lying more isovector component on the
tail of the giant dipole resonance.

\end{abstract}


\maketitle

Collective phenomena are a common feature of strongly interacting
many-body quantum systems directly linked to the relevant effective
interactions. Atomic nuclei also show collective behavior. One example
is given by the giant resonances, which have been investigated
intensively using different experimental methods, see
e.g., \cite{Hara01}. The isovector electric giant dipole resonance
(IVGDR) has been the first giant resonance to be observed in atomic nuclei.
Ever since it has been of particular interest, because collective $E1$
response is related to symmetry breaking between neutrons and protons. In
recent years, the so-called pygmy dipole resonance (PDR) \cite{Bart73,Herz97,Zilg02},
a concentration of electric dipole strength energetically below the IVGDR,
has been studied intensively in various nuclei. Within most modern
microscopic nuclear structure models, this new excitation mode is related to the
oscillation of a neutron skin against a symmetric proton-neutron core with isospin $T=0$; for an
overview see the recent review by Paar \textit{et~al.}
\cite{Paar07}. Consequently, one expects an increase of the PDR
strength approaching isotopes with extreme neutron-to-proton
ratios. Experiments on radioactive neutron-rich nuclei seem to support
this assumption \cite{Adri05,Klim07,Auma08,Gibe08,Wiel09,Wiel10}. If this
picture holds, the strength of the PDR is related to the thickness of
the neutron skin and the density dependence of the symmetry energy of
nuclear matter \cite{Klim07,Piek06}. The PDR thus permits experimental access to
these properties. However, more consistent systematic investigations
and especially more constraints on the structure of the PDR are
mandatory, such as the experiments presented in this Letter, in order
to confirm this picture.

Up to now only experiments on stable nuclei allow more detailed
investigations  of the PDR which yield additional observables in
order to understand the underlying structure of this new excitation
mode. In nuclear resonance fluorescence (NRF) experiments the
systematics of the PDR as well as its fragmentation and
fine-structure can be studied
\cite{Herz97,Piet02,Gova98,Zilg02,Hart04,Volz06,Savr08,Tonc10,Schw08}
up to the particle threshold. The mean excitation energy and the
summed transition strength $\sum{B(E1)}$$\uparrow$ (of up to 1\% of
the isovector energy weighted sum rule) show a smooth variation for
the investigated nuclei \cite{Volz06}. Different partly
contradictory microscopic calculations are able to reproduce the
mean properties of the measured photoresponse
\cite{Herz97,Litv09,Tson08,Avde09,Paar05,Colo00,lanz09}.

In order to study the structure of the PDR in greater detail
additional experiments with complementary probes are necessary.
However, owing to the high level densities and excitations of higher
multipolarity in the same energy region the experimental
investigation of the PDR using other probes than (real or virtual)
photons is very difficult. It has been shown
\cite{Poel92,Savr06b,endr09} that the \aag reaction at medium
energies provides an excellent selectivity to low-spin states
similar to NRF and, thus, represents a powerful tool to study $E1$
strength. First investigations in the N=82 isotones using this
method have indicated, in comparison to photon-scattering
experiments, an unexpected splitting of the PDR into two
well-separated groups of $J^{\pi}=1^-$ states \cite{Savr06b,
endr09}. The data suggest these two groups of $J^{\pi}=1^-$
states have different underlying structures. While recently an
isoscalar-isovector splitting of higher-lying $E1$ strength in relativistic random-phase approximation
calculations for $^{140}$Ce has been reported \cite{Paar09}, no
theoretical calculations which reproduce the experimentally observed
splitting of the PDR, are available so far.

In order to prove that the observed splitting of the PDR is not a
unique phenomenon of the N=82 isotones we have extended our
experimental studies to the Z=50 isotope \sn. In this Letter we
present the results of the \sn \aag experiment together with
calculations performed within the RQTBA and QPM models which are
able to qualitatively reproduce the observed splitting of the PDR in
\sn. Allowing for the data on \ba ~and \ce, we prove that the
splitting into two groups of states is a general feature of the PDR.
Both theoretical calculations predict a low-lying isoscalar
component of $J^{\pi}=1^-$ states which is dominated by neutron-skin
oscillations and a higher-lying more isovector group of states on
the tail of the GDR.

The experiment was performed at the AGOR cyclotron at KVI, Groningen, using a 136 MeV $\alpha$-beam and a self-supporting metallic $^{124}$Sn target with a thickness of 7.02 mg/cm$^2$. The target was enriched to 96.96 \%. For the detection of the scattered $\alpha$ particles the big-bite spectrometer (BBS) \cite{Berg95} was used, which is equipped with the EuroSuperNova (ESN) light-ion detection system \cite{Woer01}. The BBS was positioned at a central angle of 3.5$^\circ$ with an angular acceptance of 1.5$^\circ$-5.5$^\circ$. An array of seven high-purity germanium (HPGe) detectors was used to detect the emitted $\gamma$ rays in coincidence with the scattered $\alpha$ particles. For a detailed description of the setup see Ref. \cite{Savr06a}.

In the data analysis the excitation energy (deduced from the measured energy of the scattered $\alpha$ particle) is plotted versus the decay energy (obtained from the coincidently measured $\gamma$-ray energies) in a two-dimensional matrix. By applying narrow cuts on this matrix, $\gamma$ decays into different final states of $^{124}$Sn can be selected such as the ground state or the $J^{\pi}=2^+_1$ state.

\begin{figure}[h]
\includegraphics[width = 8.0 cm]{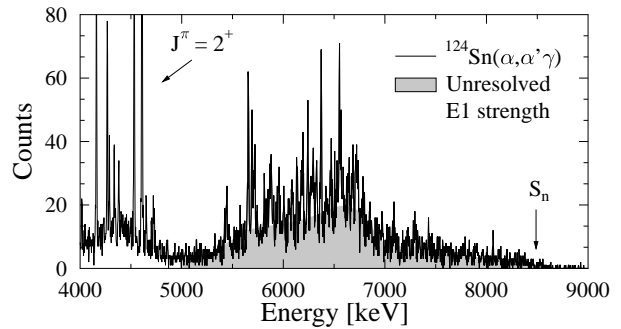}%
\caption{Summed $\gamma$-ray signals of all HPGe detectors after applying the $\alpha$-$\gamma$-coincidence condition for the ground-state decays. The PDR is clearly visible at energies above 5 MeV. \label{sn_spectrum}}
\end{figure}

The selection of decays into the ground state is very efficient to separate the $J^{\pi}=1^-$ states of the PDR from other excitations. From \krf experiments it is known that $1^-$ states decay predominantly to the ground state in contrast to states of higher multipolarity. By selecting the decays to the ground state nearly background free $\gamma$ spectra can be generated showing exclusively decays of $J^{\pi}=1^-$ states for $E_x > 5$ MeV. This demonstrates the excellent selectivity of the method. Figure \ref{sn_spectrum} shows the ground-state decays as a sum of all HPGe detectors. Each HPGe detector has an energy resolution of about 10-15 keV for $\gamma$-ray energies between 4 MeV and 9 MeV. This high resolution allows a state-by-state analysis of each single transition. Following the method presented in Ref. \cite{endr09} the multipolarity and differential $\alpha$-scattering cross section can be obtained for each individual observed excitation. Together with the known excitation energies from the NRF experiment a detailed comparison of these two complementary methods becomes feasible.

\begin{figure*}[t]
\begin{center}
\includegraphics[width = 16.0 cm]{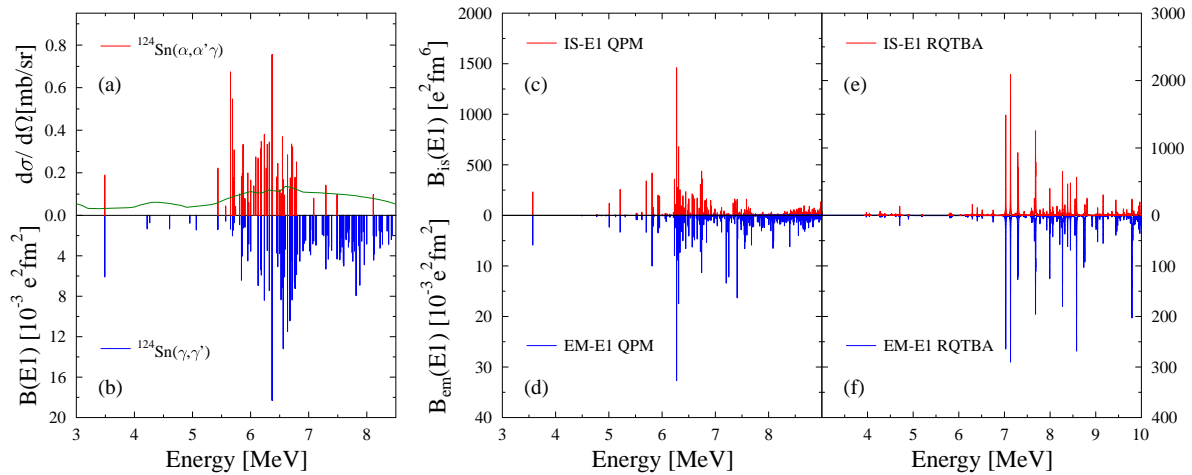}
\caption{
(a) Singles cross section for the excitation of the $J^{\pi}=1^-$ states in $^{124}$Sn obtained in the \aag coincidence experiment. The solid line shows the energy-dependent experimental sensitivity limit.
(b) $B(E1)$$\uparrow$ strength distribution measured with the \krf reaction.
The middle column shows the QPM transition probabilities in $^{124}$Sn for the
isoscalar (c) and electromagnetic (d) dipole operators.
The RQTBA strength functions in $^{124}$Sn for the
isoscalar and electromagnetic dipole
operators are shown in (e) and (f), respectively.
\label{plotall}}
\end{center}
\end{figure*}

The deduced $\alpha$-scattering cross sections for the individual states are presented in Fig.~\ref{plotall}(a). The solid line marks the sensitivity limit of the experiment. For comparison, the $B(E1)$$\uparrow$ strength distribution measured in NRF is shown in Fig.~\ref{plotall}(b). The data have been taken from Ref. \cite{Gova98} and are extended by a recent measurement of some of us where additional states at lower excitation energies have been observed at the high intensity photon setup in Darmstadt \cite{Volz06}. The structure of the spectrum shown in Fig.~\ref{sn_spectrum} suggests that strength located in the energy region of the PDR (5.5-9.0 MeV) is not entirely resolved in single transitions, especially above 7 MeV. An analysis of the angular correlation shows that also this part clearly displays a dominant dipole character.

\begin{figure}[]
\includegraphics[width = 8.5 cm]{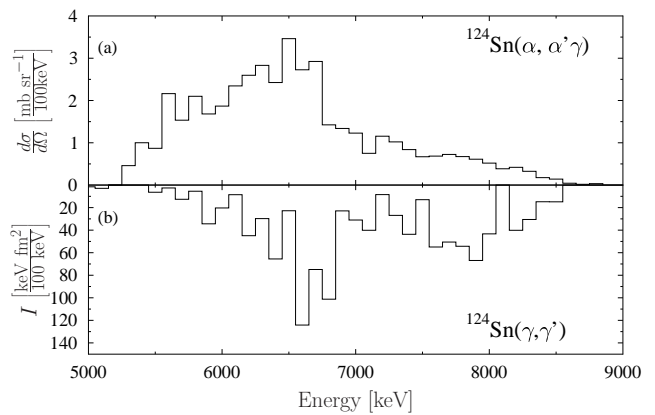}%
\caption{(a) Differential cross section obtained from the \sn \aag experiment integrated to bins with a width of 100 keV. (b) Energy integrated cross section measured in \sn \krf integrated to bins with a width of 100 keV. \label{theory_is}}
\end{figure}

In order to estimate an upper limit of this contribution we calculated the differential cross section in bins of 100 keV width for the complete shadowed part of the spectrum after subtracting the contribution of random coincidences. This integrated cross section is shown in Fig.~\ref{theory_is} in comparison to NRF data which represents a lower limit as discussed in Ref. \cite{Savr08}. We also have binned the integrated cross section deduced from the NRF experiment in 100 keV steps. However, it should be stressed that in this case only the cross sections of the single states are included since no contributions from unresolved strength are reported for the photon-scattering experiments.

The comparison of the results of the two experiments [see Figs.~\ref{plotall}(a) and \ref{plotall}(b)] shows the same remarkable behavior as already observed in $^{140}$Ce and $^{138}$Ba \cite{Savr06b, endr09}. Up to a certain excitation energy (about 6 MeV for $^{140}$Ce and $^{138}$Ba and about 6.5 MeV for \sn) all states which are known from \krf could also be observed in the \aag experiments. However, almost all higher-lying states could not be excited with the \aag reaction. This abrupt change of response to photons on the one hand and $\alpha$ particles on the other hand must be related to a structural difference between the group of $J^{\pi}=1^-$ states in the low-energy region and the group of states with higher energies.

This effect has been examined by microscopic calculations.
The $(\gamma,\gamma')$ cross sections can be directly compared
to calculated nuclear response to the electromagnetic
dipole operator $r~Y_1$.
The calculation of the $(\alpha,\alpha')$ cross sections involves
the Coulomb and nucleon-nucleon terms of the $\alpha$-particle
interaction with the target nucleus. We have checked that the former term
plays a marginal role (less than 10\%) under conditions of the present experiment.
Then, accounting for a small $q$ value of the reaction which
is about 0.33~fm$^{-1}$, the $(\alpha,\alpha ')$ cross section
is proportional with a good accuracy to the response to the
isoscalar dipole operator $r^3~Y_1$.
The spurious center-of-mass motion has been removed (see, e.g.,
\cite{Giai81} for details).

The nuclear structure part of these calculations has been performed
within the quasiparticle-phonon model (QPM) \cite{Solo92} and the
relativistic quasiparticle time-blocking approximation (RQTBA)
\cite{Litv08}, the most representative combination of the
microscopic nuclear structure models beyond QRPA. The QPM wave
functions of nuclear excited states are composed from one-, two- and
three-phonon components. The phonon spectrum is calculated within
the quasiparticle random-phase approximation (QRPA) on top of the
Woods-Saxon mean field with single-particle energies corrected to
reproduce the experimentally known single-particle levels in
neighboring odd-mass nuclei. The details of calculations are similar
to the ones in Ref. \cite{Herz97,Gova98,Savr08}. The results are
presented in Fig.~\ref{plotall}. Figure \ref{plotall}(d) shows that
the electromagnetic strength is strongly fragmented with two
pronounced peaks at about 6.3 MeV and 7.5 MeV, in good agreement
with the measured $(\gamma,\gamma ')$ data. The isoscalar response
in Fig.~\ref{plotall}(c) reveals the suppression of the strength in
the higher energy part of the spectrum, in good qualitative
agreement with the data.

The RQTBA is based on the covariant energy-density functional and
employs a fully consistent parameter-free technique (for details see
Ref. \cite{Litv08}) to account for nucleonic configurations beyond
the simplest two-quasiparticle ones. The RQTBA excited states are
built of the two-quasiparticle-phonon (2q$\otimes$phonon)
configurations, so that the model space is constructed with the
quasiparticles of the relativistic mean field and the phonons
computed within the self-consistent relativistic QRPA. Phonons of
multipolarities 2$^+$, 3$^-$, 4$^+$, 5$^-$, 6$^+$ with energies
below 10 MeV are included in the model space. The result of these
calculations is shown in Figs.~\ref{plotall}(e) and \ref{plotall}(f). Compared to the
experimental and to the QPM spectra, the structural features are
shifted by about 600 keV towards higher energies for the $E1$
electromagnetic strength and even more for the isoscalar dipole
strength. Furthermore, the obtained fragmentation is still not
sufficient. Nevertheless, the general picture demonstrates clearly
the suppression of the isoscalar dipole strength at higher energies.

\begin{figure}[]
\begin{center}
\includegraphics[width=8.0 cm]{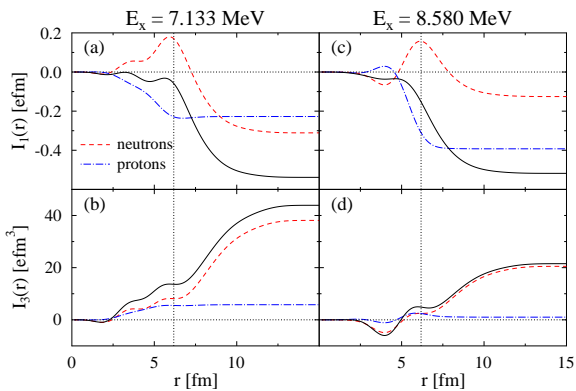}
\end{center}
\caption{Electromagnetic $I_1({\rm r})$ (top) and isoscalar $I_3({\rm r})$
(bottom) radial integrals for two particular RQTBA states - solid lines.
Contribution from neutrons (protons) is shown by dashed (dot-dashed) lines.
Vertical dotted line is plotted at the nuclear surface $R_0 = 1.24\cdot A^{1/3}$.
See text for details.}
\label{td}%
\end{figure}

The PDR pattern of the neutron skin oscillation against the $T=0$
core is important in understanding the peculiarities of the
reactions under discussion. The RQTBA states at 7.13 and 8.58~MeV
have almost equal ${\rm B_{em}(E1)}$ but differ in ${\rm
B_{is}(E1)}$ values by the factor 4. The PDR pattern is
well-developed in the first state while contribution from the
neutron skin is weaker in the second one. In general, within the
RQTBA the deviations from the PDR pattern start to develop above
$\approx$ 8 MeV. Fig.~\ref{td} plots radial integrals $I_{\eta}({\rm
r})=\int_0^{\rm r} \rho_{\rm em(is)} (r)
 ~r^{\eta +2}~dr$
where $\rho(r)$ are transition densities;
$\eta = 1$ for electromagnetic and $\eta = 3$ for isoscalar transitions.
$I_{\eta}(\infty)$ represents the corresponding transition matrix elements.
Notice the dominant role of the external part of transition densities in
both electromagnetic and isoscalar cases and its enhancement
in the latter due to the higher factor $\eta$.
While interference of the tails of neutron and proton transition
densities plays substantial role in $I_{1}(\infty)$, the
$I_{3}(\infty)$ quantity is determined by the
neutron skin.


This analysis shows that $\alpha$ particles can be expected to be more
sensitive to the surface neutron oscillation mode and less sensitive
to states showing stronger contribution of the IVGDR, as expected for
the region of the tail of the IVGDR. Therefore, the experimentally
observed splitting of the low-lying $E1$ strength which seems to be a general feature of the PDR suggests, that the
low-lying group of $1^{-}$ states actually represents the more isoscalar neutron-skin
oscillation most often associated with the interpretation of the PDR,
while the higher lying $1^{-}$ states belong to a transitional region on
the tail of the isovector GDR. However, further experimental evidence is desirable
to confirm this interpretation, as would be expected from e.g. $(p,p'\gamma)$
experiments at medium energies.

The presented results show that beside systematic
investigation of the PDR with real or virtual photons, experiments on
exotic nuclei using isoscalar and surface sensitive probes such as
$\alpha$ particles are one of the most valuable but also most challenging
demands to get a deeper understanding of this new
excitation mode in atomic nuclei.

This work was supported by the Deutsche Forschungsgemeinschaft (ZI
510/4-1 and SFB 634) and by the LOEWE program of the State of Hesse
(Helmholtz International Center for FAIR). The research has further
been supported by the EU under EURONS Contract No.
RII3-CT-2004-506065 in the 6th framework program, by the DFG cluster
of excellence Origin and Structure of the Universe, and by the
Russian Federal Education Agency Program. The authors thank P. von
Brentano, F. Iachello, H. Lenske, R. Schwengner, and A. Tonchev for
stimulating discussions.

\bibliography{prl}

\end{document}